**Modification of special relativity and local structures of gravity-free space and time**

Jian-Miin Liu*
Department of Physics, Nanjing University
Nanjing, The People's Republic of China
On leave. Present mailing address: P.O.Box 1486, Kingston, RI 02881, USA

ABSTRACT
   Besides two fundamental postulates, (i) the principle of relativity and (ii) the constancy of the one-way speed of light in all inertial frames of reference, the special theory of relativity uses the assumption about the Euclidean structure of gravity-free space and the homogeneity of gravity-free time in the usual inertial coordinate system. Introducing the so-called primed inertial coordinate system, in addition to the usual inertial coordinate system, for each inertial frame of reference, we assume the Euclidean structures of gravity-free space and time in the primed inertial coordinate system and their generalized Finslerian structures in the usual inertial coordinate system. We combine this assumption with the two postulates (i) and (ii) to modify the special theory of relativity. The modified special relativity theory involves two versions of the light speed, infinite speed c' in the primed inertial coordinate system and finite speed c in the usual inertial coordinate system. It also involves the c'-type Galilean transformation between any two primed inertial coordinate systems and the localized Lorentz transformation between any two usual inertial coordinate systems. The physical principle is: the c'-type Galilean invariance in the primed inertial coordinate system plus the transformation from the primed to the usual inertial coordinate systems. Evidently, the modified special relativity theory and the quantum mechanics theory together found a convergent and invariant quantum field theory.

I. INTRODUCTION
   The current field theory, whether classical or quantum, has been suffering from the divergence difficulties for a long time. Indeed, the infinite self-energy of an electron in quantum electrodynamics was known as early as 1929 [1], while that in classical electrodynamics was known earlier. Phenomenologically substituting several finite experimental values (particle masses and charge) for their infinities in theoretical calculations, people developed a kind of renomalization techniques to remove all divergence in some quantized field systems. However, not all quantized field systems are renormalizable, and it is hard to accept renormalizability of the kind as a basic principle to truncate those non-renormalizable field systems. Moreover, as Feynman said: "renormalization of a quantity gives up any possibility of calculating that quantity" [1], the current renormalized quantum field theory is unable to explain the mass difference between neutron and proton. Such hardron mass differences are also in groups of the π-mesons, the K-mesons, the Σ-baryons and the Ξ-baryons. The origins of the divergence difficulties lay deep within the conceptual foundations of the theory. Two foundation stones of the current quantum field theory are the special theory of relativity and the quantum mechanics theory. Since it is the case that both classical field theory and quantum field theory are plagued by the divergence difficulties, the direction to get to the roots of these difficulties seems to be in the special theory of relativity.
   According to special relativity, energy of a particle in a many-particle system is not invariant. According to special relativity, the transformation properties of total energy in the many-particle system are quite indefinite because simultaneity at distant space points in a given inertial frame of reference is no longer simultaneous in any different inertial frame of reference. These and other, such as the Lorentz non-invariance of box volume, cause inextricable difficulties in constructing invariant statistical mechanics of the many-particle system and its invariant thermodynamics in the framework of the special theory of relativity [2].

Although the special theory of relativity has been enormously successful in explaining various physical phenomena and has been tested to very high degree of precision, the above difficulties indicate that it is not a ultimate theory. Just for this, question "Is the special theory of relativity, for reasons unspecified and unknown, only an approximate symmetry of nature?" was raised [3], and the neutrino and Kaon tests to investigate possible violations of the Lorentz invariance were proposed [3,4]. A kind of modification is needed for the special theory of relativity.

In the present paper, we propose a modification for the special theory of relativity. This modification keeps the relativity principle and the constancy of the one-way speed of light but requires a change in our notion about the local structures of gravity-free space and time. The paper consists of seven sections. In Section II, we analyze experimental facts for indication how to modify the special theory of relativity. With the experimental indication, we make a study of special relativity in Section III. The special theory of relativity actually uses another assumption besides two fundamental postulates on the principle of relativity and the constancy of the one-way speed of light. But this another assumption is not an experimentally well-grounded one. We are going to make a new assumption instead of it. Since Finsler geometry and its generalization are not familiar to many people, we introduce them in Section IV before we make the new assumption in Section V. In Section VI, we combine the new assumption with the two fundamental postulates and modify the special theory of relativity. Finally, in Section VII, we draw some conclusions and make some discussions, especially, about the physical principle in the modified special relativity theory. It is argued that the modified special relativity theory and the quantum mechanics theory together found a convergent and invariant quantum field theory.

II. EXPERIMENTAL FACTS

Because of technological limitations, in the earlier experiments testing the constancy or isotropy of the light speed, light was propagated in a closed path. The favorite conclusions from these experiments are explicitly for the constancy of the speed of the round-trip propagating light, not for that of the one-way speed of light. As a result of the technological development, many experiments have been done in the manner that light propagates in a one-way. Two research groups, of Turner and Hill [5] and of Champeney et al [6], placed a $Co^{57}$ source near the rim of a standard centrifuge with an iron absorber near the axis of rotation. They used the Mossbauer effect to look for any velocity dependence of the frequency of the 14.4 KeV γ-rays as seen by the $Fe^{57}$ in the absorber. They established limits of $\Delta c/c < 2 \times 10^{-10}$ for the anisotropy in the one-way speed of light. Riis and his colleagues [7] compared the frequency of a two-photon transition in a fast atomic beam to that of a stationary absorber while the direction of the fast beam is rotated relative to the fixed stars and found the upper limit $\Delta c/c < 3.5 \times 10^{-9}$ firstly and $\Delta c/c < 2 \times 10^{-11}$ later for the anisotropy. The experiment of Krisher et al [8] was made by use of highly stable hydrogen-maser frequency standards (clocks) separated by over 21 km and connected by a ultrastable fiber optics link. The limits yielded from the experimental data are respectively $\Delta c/c < 2 \times 10^{-7}$ for linear dependency and $\Delta c/c < 2 \times 10^{-8}$ for quadratic dependency on the velocity of the Earth with respect to the cosmic microwave background.

All experimental tests of the constancy of the one-way light speed can be also interpreted as the tests of the local Lorentz invariance. Nevertheless, since local Lorentz non-invariance implies a departure from the Einstein time dilation and singles out a preferred inertial frame of reference, the experiments done by McGowan et al [9], Bailey et al [10], Kaivola et al [11], Prestage et al [12], and Krisher et al [8] can be accounted immediate testing the local Lorentz invariance. Bailey et al, Kaivola et al, and McGowan et al verified the Einstein time dilation to an accuracy of $1 \times 10^{-3}$, $4 \times 10^{-5}$ and $2.3 \times 10^{-6}$ respectively. The experiments of Prestage et al and Krisher et al are sensitive to the effects of motion of their experimental apparatus relative to a preferred inertial frame of reference and sensitive to the form of time dilation coefficient of the used hydrogen-maser clocks. However, the explicitly null results were provided in these two experiments for breakdown of the local Lorentz invariance.

Experiments clearly support the existence of the constancy of the one-way light speed, the Einstein velocity addition law and the local Lorentz invariance [5-13]. The experimental indication is: any modification of special relativity must keep the constancy of the one-way speed of light and the local Lorentz invariance.

## III. THE SPECIAL RELATIVITY THEORY

Einstein published his special theory of relativity in 1905 [14]. He derived the Lorentz transformation between any two usual inertial coordinate systems, which is the kinematical background for the physical principle of the Lorentz invariance. Two fundamental postulates stated by Einstein as the basis for his theory are (i) the principle of relativity and (ii) the constancy of the one-way speed of light in all inertial frames of reference. Besides these two fundamental postulates, the special theory of relativity also uses another assumption. This other assumption concerns the Euclidean structure of gravity-free space and the homogeneity of gravity-free time in the usual inertial coordinate system $\{x^r,t\}$, $r=1,2,3$, $x^1=x$, $x^2=y$, $x^3=z$,

$$dX^2=\delta_{rs}dx^r dx^s, \ r,s=1,2,3, \tag{1a}$$
$$dT^2=dt^2, \tag{1b}$$

everywhere and every time.

Postulates (i) and (ii) and the assumption Eqs.(1) together yield the Lorentz transformation between any two usual inertial coordinate systems [14-18]. Indeed, though the assumption Eqs.(1) was not explicitly articulated, evidently having been considered self-evident, Einstein said in 1907: "Since the propagation velocity of light in empty space is c with respect to both reference systems, the two equations, $x_1^2+y_1^2+z_1^2-c^2t_1^2=0$ and $x_2^2+y_2^2+z_2^2-c^2t_2^2=0$, must be equivalent." [17]. Leaving aside a discussion of whether postulate (i) implies the linearity of transformation between any two usual inertial coordinate systems and the reciprocity of relative velocities between any two usual inertial coordinate systems, we know that the two equivalent equations, the linearity of transformation and the reciprocity of relative velocities lead to the Lorentz transformation.

Some physicists explicitly articulated the assumption Eqs.(1) in their works on the topic. Pauli wrote: "This also implies the validity of Euclidean geometry and the homogeneous nature of space and time." [16], Fock said: "The logical foundation of these methods is, in principle, the hypothesis that Euclidean geometry is applicable to real physical space together with further assumptions, viz. that rigid bodies exist and that light travels in straight lines." [18].

Introducing the four-dimensional usual inertial coordinate system $\{x^\gamma\}$, $\gamma=1,2,3,4$, $x^4=ict$, and the Minkowskian structure of gravity-free spacetime in this coordinate system,

$$d\Sigma^2=\delta_{\alpha\beta}dx^\alpha dx^\beta, \ \alpha,\beta=1,2,3,4, \tag{2}$$

Minkowski [19] showed in 1909 that the Lorentz transformation is just a rotation in four-dimensional spacetime. He also showed how to use the four-dimensional tensor analysis for writing invariant physical laws under the Lorentz transformation. The Minkowskian structure Eq.(2) is a four-dimensional version of the assumption Eqs.(1).

## IV. GENERALIZED FINSLER GEOMETRY

Finsler geometry is a kind of generalization of Riemann geometry [20-21]. It was first suggested by Riemann as early as 1854, and studied systematically by Finsler in 1918. Its most significant work, from the viewpoint of differential geometry, has been completed by Cartan, Rund and others.

In Finsler geometry, distance ds between two neighboring points $x^k$ and $x^k+dx^k$, $k=1,2,\cdots,n$ is defined by a scale function

$$ds=F(x^1,x^2,\cdots,x^n,dx^1,dx^2,\cdots dx^n)$$

or simply

$$ds=F(x^k,dx^k), \ k=1,2,\cdots,n, \tag{3}$$

which depends on directional variables $dx^k$ as well as coordinate variables $x^k$. Apart from several routine conditions like smoothness, the main constraint imposed on this scale function is that it is positively homogeneous of degree one in $dx^k$,

$$F(x^k,\lambda dx^k)=\lambda F(x^k,dx^k) \ \text{for} \ \lambda>0. \tag{4}$$

Introducing a set of equations

$$g_{ij}(x^k,dx^k)=\partial^2 F^2(x^k,dx^k)/2\partial dx^i \partial dx^j, \ i,j=1,2,\cdots,n, \tag{5}$$

we can represent Finsler geometry in terms of

$$ds^2=g_{ij}(x^k,dx^k)dx^i dx^j, \tag{6}$$

where $g_{ij}(x^k, dx^k)$ is called the Finslerian metric tensor induced from scale function $F(x^k, dx^k)$. The Finslerian metric tensor is symmetric in its subscripts and all its components are positively homogeneous of degree zero in $dx^k$,

$$g_{ij}(x^k, dx^k) = g_{ji}(x^k, dx^k), \tag{7a}$$

$$g_{ij}(x^k, \lambda dx^k) = g_{ij}(x^k, dx^k) \text{ for } \lambda > 0. \tag{7b}$$

We can define the so-called generalized Finsler geometry by omitting Eq.(3) as a definition of ds and instead taking Eq.(6) as a definition of $ds^2$, where the given metric tensor $g_{ij}(x^k, dx^k)$ satisfies Eqs.(7) and other routine conditions.

The generalized Finsler geometry is so-named because a Finsler geometry must be a generalized Finsler geometry but the inverse statement is not valid, in other words, a generalized Finsler geometry is not necessarily a Finsler geometry [21].

A generalized Finsler geometry with a given metric tensor $g_{ij}(x^k, dx^k)$ is also a Finsler geometry when and only when we can find a scale function $F(x^k, dx^k)$ by solving the set of equations, Eqs.(5), such that this function, as a solution to the set of equations Eqs.(5), is positively homogeneous of degree one in $dx^k$. In the case, $F^2(x^k, dx^k)$ is positively homogeneous of degree two in $dx^k$. The following equation thus holds due to Euler's theorem on homogeneous functions,

$$2F^2(x^k, dx^k) = dx^i [\partial F^2(x^k, dx^k) / \partial dx^i]. \tag{8}$$

Iterating Eq.(8), we find

$$4F^2(x^k, dx^k) = dx^j \delta_{ij} [\partial F^2(x^k, dx^k) / \partial dx^i]$$
$$+ dx^j dx^i [\partial^2 F^2(x^k, dx^k) / \partial dx^j \partial dx^i]. \tag{9}$$

It then follows from use of Eq.(8) that

$$F^2(x^k, dx^k) = dx^j dx^i [\partial^2 F^2(x^k, dx^k) / 2 \partial dx^j \partial dx^i]. \tag{10}$$

Eq.(10) combines with Eqs.(5) and (6) to yield Eq.(3).

Like Finsler geometry, generalized Finsler geometry can be endowed with the Cartan connection.

## V. A NEW ASSUMPTION ON LOCAL STRUCTURES OF SPACE AND TIME

Conceptually, the principle of relativity implies that there exists a class of the equivalent inertial frames of reference, any one of which moves with a non-zero constant velocity relative to any other. Einstein wrote: "in a given inertial frame of reference the coordinates mean the results of certain measurements with rigid (motionless) rods, a clock at rest relative to the inertial frame of reference defines a local time, and the local time at all points of space, indicated by synchronized clocks and taken together, give the time of this inertial frame of reference."[15]. As defined by Einstein, each of the inertial frames of reference is supplied with motionless, rigid unit rods of equal length and motionless, synchronized clocks of equal running rate. Then, in each inertial frame of reference, an observer can employ his own motionless-rigid rods and motionless-synchronized clocks in the so-called "motionless-rigid rod and motionless-synchronized clock" measurement method to measure space and time intervals. By using this "motionless-rigid rod and motionless-synchronized clock" measurement method, the observer in each inertial frame of reference can set up his own usual inertial coordinate system. Postulate (ii) means that the speed of light is measured to be the same constant c in every such usual inertial coordinate system.

The "motionless-rigid rod and motionless-synchronized clock" measurement method is not the only one that each inertial frame of reference has. We imagine, for each inertial frame of reference, other measurement methods that are different from the "motionless-rigid rod and motionless-synchronized clock" measurement method. By taking these other measurement methods, an observer in each inertial frame of reference can set up other inertial coordinate systems, just as well as he can set up his usual inertial coordinate system by taking the "motionless-rigid rod and motionless-synchronized clock" measurement method. We call these other inertial coordinate systems the unusual inertial coordinate systems.

The conventional belief in flatness of gravity-free space and time is natural. But question is, in which inertial coordinate system the gravity-free space and time directly display their flatness. The special theory of relativity recognizes the usual inertial coordinate system, as shown in Eqs.(1). Making a different choice, we take one of the unusual inertial coordinate systems, say $\{x'^r, t'\}$, r=1,2,3, the primed inertial coordinate system. We assume that gravity-free space and time possess the flat metric structures in the primed inertial coordinate system and the following generalized Finslerian structures in the usual inertial coordinate system,

$$dX^2 = \delta_{rs}dx'^r dx'^s = g_{rs}(y)dx^r dx^s, \quad r,s=1,2,3, \tag{11a}$$
$$dT^2 = dt'^2 = g(y)dt^2, \tag{11b}$$
$$g_{rs}(y) = K^2(y)\delta_{rs}, \tag{11c}$$
$$g(y) = (1 - y^2/c^2), \tag{11d}$$
$$K(y) = \frac{c}{2y} \ln \frac{c+y}{c-y} (1 - y^2/c^2)^{1/2}, \tag{11e}$$

where $y = (y^s y^s)^{1/2}$, $y^s = dx^s/dt$, $s=1,2,3$.

Two metric tensors $g_{rs}(y)$ and $g(y)$ depend only on directional variables $y^s$, $s=1,2,3$, and become flat when and only when y approaches zero.

## VI. THE MODIFIED SPECIAL RELATIVITY THEORY

We now modify the special theory of relativity by combining the alternative assumption Eqs.(11), instead of the assumption Eqs.(1), with the two postulates (i) and (ii).

If we define a new type of velocity, $y'^s = dx'^s/dt'$, $s=1,2,3$, in the primed inertial coordinate system and keep the well-defined Newtonian velocity in the usual inertial coordinate system, we find from the assumption Eqs.(11)

$$y'^s = [\frac{c}{2y} \ln \frac{c+y}{c-y}] y^s, \quad s=1,2,3, \tag{12}$$

and

$$y' = \frac{c}{2} \ln \frac{c+y}{c-y}, \tag{13}$$

where $y' = (y'^s y'^s)^{1/2}$, $s=1,2,3$. It is understood that two different measurement methods can be applied to a motion when the motion is observed in an inertial frame of reference, one being the "motionless-rigid rod and motionless-synchronized clock" measurement method, the other one being associated with the primed inertial coordinate system. As a result, two different velocities are obtained, primed velocity $y'^s$ (of the new type) and Newtonian velocity $y^s$. Velocities $y'^s$ and $y^s$ are two versions of the motion, obtained via two different measurement methods used in the inertial frame of reference. They are related by Eqs.(12) and (13). The Galilean addition among primed velocities links up with the Einstein addition among usual velocities [22]. This statement can be easily seen in the one-dimensional case that

$$y'_2 = y'_1 - u' = (c/2)\ln[(c+y_1)/(c-y_1)] - (c/2)\ln[(c+u)/(c-u)]$$

and

$$y'_2 = (c/2)\ln[(c+y_2)/(c-y_2)]$$

imply

$$y_2 = (y_1 - u)/(1 - y_1 u/c^2).$$

In Eq.(13), as y goes to c, we get an infinite primed speed,

$$c' = \lim_{y \to c} \frac{c}{2} \ln \frac{c+y}{c-y}. \tag{14}$$

Speed c' is invariant in the primed inertial coordinate systems simply because speed c is invariant in the usual inertial coordinate systems. Speed c' is really a new version of the light speed, its version in the primed inertial coordinate systems.

Let IFR1 and IFR2 be two inertial frames of reference, where IFR2 moves with non-zero Newtonian velocity $u^s$, $s=1,2,3$, relative to IFR1. IFR1 and IFR2 can use their own "motionless-rigid rod and motionless-synchronized clock" measurement methods and set up their own usual inertial coordinate systems $\{x^r_m, t_m\}$, $m=1,2$. They can also set up their own primed inertial coordinate systems, $\{x'^r_m, t'_m\}$, $m=1,2$. Since the propagation velocity of light is c' in both $\{x'^r_1, t'_1\}$ and $\{x'^r_2, t'_2\}$, we have two equivalent equations,

$$\delta_{rs}dx'^r_1 dx'^s_1 - c'^2(dt'_1)^2 = 0, \tag{15a}$$
$$\delta_{rs}dx'^r_2 dx'^s_2 - c'^2(dt'_2)^2 = 0. \tag{15b}$$

Using Eqs.(11) with y=c, we have further two equivalent equations,

$$\delta_{rs}dx_1^r dx_1^s - c^2(dt_1)^2 = 0, \quad (16a)$$
$$\delta_{rs}dx_2^r dx_2^s - c^2(dt_2)^2 = 0, \quad (16b)$$

because $c^2 K^2(c) = c'^2 g(c)$, where $K(c) = \lim_{y \to c} K(y)$, $g(c) = \lim_{y \to c} g(y)$.

Two equivalent equations Eqs.(15), the linearity of transformation between two $\{x'^r_m, t'_m\}$, the reciprocity of relative velocities between two $\{x'^r_m, t'_m\}$, and the flat structures of gravity-free space and time in two $\{x'^r_m, t'_m\}$ will lead to the c'-type Galilean transformation between two primed inertial coordinate systems $\{x'^r_m, t'_m\}$, m=1,2, under which speed c' is invariant. Two equivalent equations Eqs.(16), the linearity of transformation between two $\{x^r_m, t_m\}$, and the reciprocity of relative velocities between two $\{x^r_m, t_m\}$ will lead to the localized Lorentz transformation between two usual inertial coordinate systems $\{x^r_m, t_m\}$, m=1,2,

$$dx_2^r = (dx_1^r - u^r dt_1) + (\gamma-1)u^r u^s (dx_1^s - u^s dt_1)/u^2, \quad r,s=1,2,3, \quad (17a)$$
$$dt_2 = \gamma(dt_1 - u^k dx_1^k/c^2), \quad k=1,2,3, \quad (17b)$$

where

$$\gamma = 1/(1-u^2/c^2)^{1/2} \quad (18)$$

and $u = (u^s u^s)^{1/2}$, s=1,2,3.

In the modified special relativity theory, the c'-type Galilean transformation stands between any two primed inertial coordinate systems, while the localized Lorentz transformation between two corresponding usual inertial coordinate systems. Substituting the assumption Eqs.(11) for the assumption Eqs.(1) does not spoil the localized Lorentz transformation between any two usual inertial coordinate systems. The modified special relativity theory keeps the constancy of the one-way speed of light and also the local Lorentz invariance in the usual inertial coordinate system, as well as, the c'-type Galilean invariance in the primed inertial coordinate system.

## VII. CONCLUDING REMARKS AND DISCUSSIONS

In the context, we have introduced the primed inertial coordinate system, in addition to the usual inertial coordinate system, for each inertial frame of reference. We have assumed the flat structures of gravity-free space and time in the primed inertial coordinate system and their generalized Finslerian structures in the usual inertial coordinate system. Combining this assumption with the two postulates (i) and (ii), we have modified the special theory of relativity. The modified special relativity theory involves two versions of the light speed, infinite speed c' in the primed inertial coordinate system and finite speed c in the usual inertial coordinate system. It also involves the c'-type Galilean transformation between any two primed inertial coordinate systems and the localized Lorentz transformation between any two usual inertial coordinate systems.

Consequently, we shall have to make a change in the physical principle. In the special theory of relativity, the physical principle is the Lorentz invariance: All physical laws keep their forms with respect to the Lorentz transformation in the usual inertial coordinate system. In the modified special relativity theory, since it is the "motionless-rigid rod and motionless-synchronized clock" measurement method that we use in our experiments, the physical principle is: The c'-type Galilean invariance in the primed inertial coordinate system plus the transformation from the primed inertial coordinate system to the usual inertial coordinate system. We write all physical laws in the c'-type Galilean-invariant form in the primed inertial coordinate system and do all calculations in the c'-type Galilean-invariant manner. We finally transform all calculation results from the primed inertial coordinate system to the usual inertial coordinate system, and compare them to experimental facts in the usual inertial coordinate system.

This new physical principle has been applied to reform of mechanics and field theory. Evidently, relativistic mechanics is still valid in the usual inertial coordinate system, while field theory is freshened. Any field system can be quantized in the primed inertial coordinate system with use of canonical quantization method. Evidently, any quantized field system will undergo a unitary transformation as it is transformed from the primed to the usual inertial coordinate systems. Particle size defined in the primed inertial coordinate system is an invariant quantity. It can be quite involved in the c'-type Galilean-invariant calculations. In this way, the modified special relativity theory and the quantum mechanics theory together will found a convergent and invariant quantum field theory. Readers who are interested in detailed derivations for these may refer to Ref.[23].

ACKNOWLEDGMENT

The author greatly appreciates the teachings of Prof. Wo-Te Shen. The author thanks Prof. Mark Y.-J. Mott, Dr. Allen E. Baumann and Dr. C. Whitney for helpful suggestions.